\documentclass[a4paper,12pt]{article}
\begin{document}
\begin{center}
\large \bf COULOMB TRANSITION MATRIX WITH FRACTIONAL VALUES OF INTERACTION PARAMETER \\[.4in]
\end{center}

{\sf V.F. Kharchenko}\\[.1in]
{\footnotesize Bogolyubov Institute for Theoretical Physics,  
National Academy of Sciences \\of Ukraine, UA - 03143, Kyiv, Ukraine \\ [.1in]
E-mail: vkharchenko@bitp.kiev.ua} \\[.1in]

\noindent \small{ {\sf Abstract }  \\
{\small Leaning upon the specific Fock symmetry of the Coulomb interaction
potential in the four-dimensional momentum space we perform the analytical 
solution of the Lippman-Schwinger equation for the Coulomb transition 
matrix in the case of negative energy at fraction values of the interaction 
parameter. Analytical expressions for the three-dimensional and partial 
Coulomb transition matrix with simplest factional values of the interaction 
parameter are obtained. } \\ [.05in]

\noindent {\sf 1. Introduction} \\ 

\noindent The Faddeev integral equations describing three-body system contain the 
two-body Coulomb transition matrix off-energy shell. The Coulomb transition 
matrix, like to the Coulomb Green function, contains all information about 
two-particle system.
   In this paper, using the symmetry of the Coulomb system in four-dimensional 
   Euklidean space first described by Fock [1], the analytical solution of
the Lippman-Schwinger equation for the Coulomb transition matrix at integer 
and half-integral values of the interaction parameter.
   In the section 2 the basic formulas for the three-dimensional Coulomb 
transition matrix are given.
   The section 3 is devoted to the analytic expressions for the Coulomb 
transition.matrix in the case  of repulsive and attractive Coulomb interaction 
at fractional values of the Coulomb parameter.
   The section 4 is devoted to conclusions.\\

\noindent {\sf 2. The Coulomb transition matrix}\\ 

   It is known that two-particle transition operator in the three-dimensional 
space satisfies the inhomogeneous Lippman-Schwinger integral equation 
\begin{equation}
t^C(E) = v^C + v^C g_0(E)t^C(E)                            
\end{equation}                                            
where $v^C$ is operator of the Coulomb interaction, $g_0(E)$ is the free Green 
function,$E$ is the energy of the relative motion.of the two particles.
   In the momentum space the matrix of the potential Coulomb interation of 
the partiles 1 and 2 hss the form
\begin{equation}
<\vec{k}|v^C|\vec{k'}>=\frac{4 \pi q_1 q_2}{|\vec{k}-\vec{k'}|^2}       
\end{equation} 
$q_i$ is the charge of the particle $i$, $\vec{k}$ and $\vec{k'}$ are the variable 
of the relative motion of particles. In this paper we restrict our consideration 
to the two-body system with the negative energy
\begin{equation}
E=-\frac{\hbar^2\kappa^2}{2\mu},                      
\end{equation}
where $\mu$ is the reduced mass of the particles, $\hbar$ is the Planc constant.
	
  The solution of the integral equation for the three-dimensional transition 
matrix $<\vec{k}|t^C(E)|\vec{k'}>$ at the negative energy may be written using 
the Fock method of the stereographic projection at the momentum space onto the 
four-dimensional sphere of the unit radius [1] in the form [2]

\begin{equation}
<\vec{k}|t^C(E)|\vec{k'}>= \frac{2\pi q_1 q_2 \eta}{kk'}
[\frac{1}{sin^2\frac{\omega}{2}}-\frac{4\gamma}{\sin\omega}
\sum\limits_{n=1}^\infty\frac{\sin{n\omega}}{n+\gamma}],      
\end{equation}

where
\begin{equation} 
\eta=\frac{2 \kappa^2 kk'}{(k^2+\kappa^2)(k'^2+\kappa^2)},    
\end{equation}

$\gamma$ is nondimensional parameter of the Coulomb interaction (the Sommerfeld 
parameter)

\begin{equation}
\gamma=\frac{\mu q_1 q_2}{\hbar^2 \kappa}                     
\end{equation}

and the variable $\omega$ is the angle between two vectors in the fourdimensional 
Fock space [1]

\begin{equation}  
\sin^2\frac{\omega}{2}=\frac{\kappa^2|\vec{k}-\vec{k'}|^2}{(k^2+\kappa^2)
(k'^2+\kappa^2)}, 0\leq\omega<\pi.                                                 
\end{equation}  
     \\

\noindent {\sf 3. The case with half integer Coulomb parameter}\\

 In the simplest case of two charged particles with the Coulomb parameters 
$\gamma=1/2$ and $\gamma=-1/2$ after suming with the use of the formula (3)
\begin{equation}
\sum\limits_{n=1}^\infty\frac{\sin{n\omega}}{n\pm 1/2}=
\frac{\pi}{2} \cos\frac{\omega}{2} \pm\sin\frac{\omega}{2} \ln|\tan\frac{\omega}{4}|,  
\end{equation}                                              
we find such expression for the Coulomb transition matrix [4]
\begin{equation}
<\vec{k}|t^C(E)|\vec{k'}>_{\gamma=\pm 1/2} = \frac{2\pi q_1 q_2 \eta}{kk'}
[\frac{1}{\sin^2\frac{\omega}{2}} -  \frac{\pi}{2\sin\frac{\omega}{2}} 
\mp\frac{1}{\cos\frac{\omega}{2}}\ln|\tan\frac{\omega}{4}|]                    
\end{equation}
Note that the result (9) follows immediately also from the Schwinger 
formula [4]
\begin{equation}
<\vec{k}|t^C(E)|\vec{k'}>=\frac{2\pi q_1 q_2 \eta}{kk'}
[\frac{1}{\sin^2\frac{\omega}{2}}-4\gamma\int_0^1 d\rho 
\frac{\rho^\gamma}{\rho^2 -2\rho \cos\omega + 1}],              
\end{equation}
just as from the formula for the Coulomb transition matrix obtained with 
explicit separation of the singularities in the transfer momentum
and the energy in [5]

\begin{eqnarray}
<\vec{k}|t^C(E)|\vec{k'}>& = &\frac{2\pi q_1 q_2 \eta}{kk'}[\frac{1}{\sin^2\frac{\omega}{2}}(\pi\gamma \cos{\gamma\omega}+
\gamma\sin{2\gamma\omega} \ln|\sin\frac{\omega}{2}| \nonumber \\ 
&  & - 2\pi\gamma c(\gamma)\cot{\gamma\pi}\sin{\gamma\omega} - 
\gamma \cos{\gamma\omega}\ x_{\gamma}(\omega) - 2\gamma^2 \sin{\gamma\omega} y_\gamma(\omega))],\label{line2}
\end{eqnarray}

where

\begin{equation}
 \begin{array}{rcl}
x_{\gamma} (\omega)& = &\int_{0}^{\omega} d\varphi \sin{\gamma\varphi}\cot\frac{\varphi}{2}, \\
y_{\gamma}(\omega)& = &\int_{\omega}^{\pi}d\varphi \sin{\gamma\varphi} \ln|\sin\frac{\omega}{2}| \\       
c(\gamma)& = &\frac{1}{2}[1 - \frac{1}{\pi}x_{\gamma}(\pi)].        
\end{array}
  \end{equation}
that at $\gamma=\pm \frac{1}{2}$ takes the form 
\begin{eqnarray} 
x_{\pm \frac{1}{2}}(\omega)& = &\pm 2\sin\frac{\omega}{2},\nonumber\\
y_{\pm \frac{1}{2}}(\omega)& = &\pm 2\cos\frac{\omega}{2}[\ln|\sin\frac{\omega}{2}| - 1]\pm2\ln|\cot\frac{\omega}{4}|,\label{line2}\\
c(\pm \frac{1}{2})& = &\frac{1}{2}\mp \frac{1}{\pi}. \nonumber
\end{eqnarray}

Analogously, using the formula for                                                
$\sum\limits_{k=1}^\infty\frac{1}{k+\frac{n}{m}\sin{kr}}$  
providing in [3], we obtain
\begin{eqnarray}
\sum\limits_{k=1}^\infty\frac{\sin{kx}}{k+n/m}& = &-\frac{1}{2 m} 
\sum\limits_{k=0}^{m-1}[x+(2 k-m)\pi]\cos[(x+2 k\pi)\frac{n}{m}]+ \nonumber \\
& &\sum\limits_{k=0}^{m-1}\sin(x+2 k\pi)\frac{n}{m} \ln(2|\sin\frac{x+2 k\pi}{2 m}|), \label{line2}       
\end {eqnarray}
if $\frac{n}{m}\leq1$, and 

\begin{eqnarray}
\sum\limits_{k=0}^\infty\frac{\sin{kx}}{k+n/m}& = &-\frac{1}{2m}\sum\limits_{k=0}^{m-1}
[(2 k-m)\pi+x]\cos[(x+2 k \pi)\frac{n}{m}]+ \nonumber\\
 &  &\sum\limits_{k=0}^{m-1}\sin[(x+2 k \pi)\frac{n}{m}] \ln(2|\sin\frac{x+2 k\pi}{m}|)- 
\sum\limits_{k=1}^{(n-1)/m}\frac{\sin{kx}}{k-\frac{n}{m}}, \label{line2}                                   
\end{eqnarray}

if $\frac{n}{m}>1$, we obtain the formulas for the Coulomb t-matrix in 
the general case  $\gamma=\pm\frac{n}{m}$.
(Note that in (15) we use the notation [(n-1)/m] for the integer part 
of the upper limit of the sum).
Suitably we find
\begin{equation}
<\vec{k}|t^C(E)|\vec{k'}>_{\gamma=\pm\frac{3}{2}}=\frac{2\pi q_1 q_2{\eta}}{k k'}
[\frac{1}{\sin^2\frac{\omega}{2}}\mp 3\pi \frac{\cos(\frac{3}{2}\omega)}{\sin\omega}  
-\frac{6\sin\frac{3}{2}\omega}{\sin{\omega}}\ln|\tan\frac{\omega}{2}|-12],
\end{equation}

\begin{eqnarray}
<\vec{k}|t^C(E)|\vec{k'}>_{\gamma=\pm\frac{5}{2}}=\frac{2\pi q_1 q_2{\eta}}{k k'}
[\frac{1}{\sin^2\frac{\omega}{2}}\mp 5\pi \frac{\cos\frac{5}{2}\omega}{\sin\omega} - \nonumber\\ 
\frac{10\sin\frac{5}{2}\omega}{\sin{\omega}}\ln|\tan\frac{\omega}{4}|- 
40\cos\omega-\frac{20}{3}] \label{line 2}   
\end{eqnarray}         

\begin{eqnarray} 
<\vec{k}|t^C(E)|\vec{k'}>_{\gamma=\pm\frac{7}{2}}=\frac{2\pi q_1 q_2{\eta}}{k k'}
[\frac{1}{\sin^2\frac{\omega}{2}} \mp 7\pi \frac{\cos\frac{7}{2}\omega}{\sin\omega} - \nonumber\\  
\frac{14\sin\frac{7}{2}\omega}{\sin{\omega}}\ln|\tan\frac{\omega}{4}|-
112\cos^2{\omega}-\frac{56}{3}\cos\omega+ \frac{112}{5}]. \label{line 2}
\end{eqnarray}
Using the formula (14) for m/n<1, we find the Coulomb transition matrix for  
the values $\gamma = \pm \frac{1}{3}$ and $\gamma = \pm \frac{1}{4}$ :        
%(19)%
\begin{eqnarray} 
<\vec{k}|t^C(E)|\vec{k'}>_{\gamma = \pm\frac{1}{3}} = \frac{2\pi q_1 q_2{\eta}}{k k'}
[\frac{1}{\sin^2\frac{\omega}{2}} \mp \frac{2\pi}{3}\frac{\cos\frac{\omega}{3}-
\frac{1}{\sqrt{3}}\sin\frac{\omega}{3}}{\sin\omega}+ \nonumber\\ 
\frac{2\sin\frac{\omega}{3}}{3\sin{\omega}}\ln|\frac{\sin\frac{\omega}{6}-
3\cos\frac{\omega}{6}}{4\sin^2\frac{\omega}{6}}|- \frac{2\cos\frac{\omega}{3}}
{3\sqrt{3}\sin{\omega}}\ln|\frac{\tan\frac{\omega}{6}+\sqrt{3}}
{\tan\frac{\omega}{6}-\sqrt{3}}|] \label{line2}                                                
\end{eqnarray} 

\begin{eqnarray} 
<\vec{k}|t^C(E)|\vec{k'}>_{\gamma=\pm\frac{1}{4}}= \frac{2\pi q_1 q_2{\eta}}{k k'}
[\frac{1}{\sin^2\frac{\omega}{2}}\mp \pi/2 \frac{\cos\frac{\omega}{4}- 
\sin\frac{\omega}{4}}{\sin\omega}- \nonumber\\
\frac{\sin\frac{\omega}{4}}{\sin{\omega}}
\ln|\tan\frac{\omega}{8}|-\frac{\cos{\frac{\omega}{4}}}{\sin\omega}\ln|\frac{1+\tan\frac{\omega}{8}}
{1-\tan\frac{\omega}{8}}|] \label{line2}                                                         
\end{eqnarray} 

\noindent {\sf 4. Conclusions}\\ 

    In this paper firstly established the possibility of the solution of the 
equation for the three-dimensional and partial Coulomb transition matrices 
using the specific symmetry of the Coulomb system in the Fock fourdimensional 
space for semi-integral and fractional values of the Coulomb parameter. Simple 
analytical expressions for partial and three-dimensional Coulomb transition 
matrices in the cases of the attractive and repulsive Coulomb interactions.
The developed method is used for the semi-integral negative and fractional 
values of the Coulomb parameters. These results can be used in the case of 
the formulation and solution of the equations  for the few-body systems with 
the charged particles. In this way this approach permits to gain new information 
about the fundamental nuclear interaction.  \\

\noindent {\footnotesize {\sf References} 
\vspace*{.1in}
\begin{itemize} 
\setlength{\baselineskip}{.1in}
\item[{\tt [1]}] V.A. Fock, Z. Phys. 98 (1935) 145 - 154. \\
\item[{\tt [2]}] V.F. Bratsev, E.D> Trifonov, Vestnik of Leningrad University 16 (1968) 36 - 39. \\
\item[{\tt [3]}] A.P. Prudnikov, Yu.A. Bychkov, O.I. Marichev, Integrals and Series, Moscow, 
   Nauka,1981, p.729, formulas 5.4.3.8 and 5.4.3.13. \\ Note that in the formula 
   5.4.3.13 in the second term the coeficient 1/4 is omitted. \\ 
\item[{\tt [4]}] J. Schwinger, J. Math, Phys. 5 (1964) 1606 - 1608. \\
\item[{\tt [5]}] S.A. Shadchin, V.F. Kharchenko, J. Phys. B16 (1983) 1319 - 1322. \\
\end{itemize}} 

\end{document}